\documentclass[aps,pra,showpacs,twocolumn,amsmath,preprintnumbers,nofootinbib,secnumarabic,a4paper]{revtex4} 
\usepackage{epsfig,verbatim,hyperref}
\usepackage{amssymb,amsfonts}
\usepackage{graphicx}

\usepackage[caption=false]{subfig}
\def\sc{1.02} 

\def\jrn#1#2#3#4#5#6{\textit{#3} \textbf{#4}, #5 (#6).} \def\boo#1#2#3#4#5#6{\textit{#2} (#3, #4, #5).}    \def\andd{ and } \def\andt{ and } \def\eq{Eq.\,} \def\eqs{Eqs.\,} \def\Ref{Ref.\,}  

\def\boldsymbol#1{#1}
\def\text#1{{\mbox{#1}}}

\def\eqref#1{(\ref{#1})}

\def\scn#1#2{\section{#1}\lb{#2}} \def\sscn#1#2{\subsection{#1}\lb{#2}} 

\def\bfl{\begin{flushleft}}
\def\efl{\end{flushleft}}
\def\bfr{\begin{flushright}}
\def\efr{\end{flushright}}
\def\bc{\begin{center}}
\def\ec{\end{center}}
\def\be{\begin{equation}}
\def\ee{\end{equation}}
\def\bse{\begin{subequations}}
\def\ese{\end{subequations}}
\def\ba{\begin{eqnarray}}
\def\ea{\end{eqnarray}}
\def\baa#1{\begin{array}{#1}}
\def\eaa{\end{array}}
\def\bw{\begin{widetext}}
\def\ew{\end{widetext}}
\def\nn{\nonumber }
\def\lb#1{\label{#1}}
\def\bit{\begin{itemize}}
\def\eit{\end{itemize}}
\def\bco{}
\def\bcs{\begin{cases}}
\def\ecs{\end{cases}}

\def\schrod{Schr\"odinger}

\def\Der#1#2{\frac{\drm #1}{\drm #2}}
\def\pDer#1#2{\frac{\partial #1}{\partial #2}}

\def\vena{\boldsymbol{\nabla}}

\def\nc0{\tilde b_0}

\def\vol{{\cal V}}
\def\vol{V}

\def\drm{d}
\def\dvol{\drm\vol}

\def\dn{\rho}

\def\text#1{{\rm #1}}

\begin{document}

\preprint{\small Int. J. Mod. Phys. B \textbf{35}, 2150229 (2021) 
\quad 
[\href{https://doi.org/10.1142/S0217979221502295}{DOI: 10.1142/S0217979221502295}]
}

\title{
Acoustic oscillations in cigar-shaped logarithmic Bose-Einstein condensate in the Thomas-Fermi approximation
}

\author{Konstantin G. Zloshchastiev}
\email{https://bit.do/kgz}
\affiliation{Institute of Systems Science, Durban University of Technology, P.O. Box 1334, 
Durban 4000, South Africa\\
kostiantynz@dut.ac.za\\
kostya@u.nus.edu\\~~}


\begin{abstract} 
We consider the dynamical properties of
density fluctuations in the cigar-shaped Bose-Einstein condensate described by the logarithmic wave equation
with a constant nonlinear coupling
by using the Thomas-Fermi and linear approximations.
It is shown that the propagation of small density fluctuations along the long axis of a condensed lump in a strongly anisotropic trap is essentially one-dimensional,
while the trapping potential can be disregarded in the linear regime.
Depending on the sign of nonlinear coupling, 
the fluctuations either take the form of translationally symmetric pulses 
and standing waves,
or become oscillations with varying amplitudes.
We also study the condensate in an axial harmonic trap,
by using elasticity theory's notions. 
Linear particle
density and energy also behave differently depending on the nonlinear coupling's value.
If it is negative, the density monotonously grows along with lump's radius,
while energy is a monotonous function of density.
For the positive coupling, the density is bound from above,
whereas energy grows monotonously as a function of density until it reaches its global maximum.
\end{abstract}

\date{received: 2 June 2021 [WSPC]}

\pacs{03.75.Kk, 05.30.Jp, 67.25.dt, 67.85.-d
\\ \textbf{Keywords}: quantum Bose liquid, Bose-Einstein condensate, logarithmic condensate, speed of sound, sound propagation, Thomas-Fermi approximation, cold gases.
}

\maketitle

\scn{Introduction}{s:in}
In the now classical experiments where magnetically trapped alkali atoms 
were
cooled
down to nanokelvin temperatures, a steep narrowing 
of the velocity and density distribution profiles was demonstrated \cite{aem95,bst95},
which was attributed to the occurrence of Bose-Einstein condensation (BEC),
earlier known to exist in liquified noble gases \cite{lo38}.
From a theoretical point of view,
Bose-Einstein condensates fall into a wide category of 
quantum Bose liquids.
These are a special kind of quantum matter, allowing description in terms of hydrodynamic 
degrees of freedom \cite{ttbook,psbook}. 
This kind includes media formed, not only by integer-spin particles, 
but also by the ``bosonized'' combinations of fermions.
The standard theory and classification of quantum Bose liquids is far from its completion;
due to the obvious complexity of this matter, and the large number of phenomena involved,
including not only many-body atom-atom interactions, but also vacuum effects.

Apart from the well-known Gross-Pitaevskii (GP) model \cite{gr61,pi61}, which 
belongs to a class of perturbatively defined BEC models with polynomial nonlinearity \cite{ks92};
a conceptually different model exists, defined by the wave equation with 
non-polynomial (logarithmic) nonlinearity \cite{z11appb,az11}.
One can demonstrate that  logarithmic nonlinearity should universally occur in systems
with the following properties:
(i) allowing a 
hydrodynamic description in terms of the fluid wavefunction (cf. \Ref \cite{ry99}),
and (ii) their particles' characteristic interaction potentials being 
substantially larger than kinetic energies \cite{z18zna}. 

Profound applications of the logarithmic model were found in various 
physical systems \cite{z12eb,z19ijmpb,sz19,z20un1,z21ltp,lz21cs},
and
a relation between logarithmic models and quantum information entropy should be mentioned as well \cite{az11}.
While the Gross-Pitaevskii model
was historically first in describing Bose-Einstein 
condensation of diluted systems, by using a two-body approximation;
the logarithmic model significantly advances our 
understanding of nonperturbative quantum effects in quantum fluids,
especially when used with polynomial BEC models, cf. \Ref \cite{sz19}.

In this paper,
we study the dynamics of density fluctuations in a cigar-shaped, 
or elongated,
logarithmic Bose-Einstein condensate.
In section \ref{s:mod}, we give a brief description of the logarithmic BEC model,
and enumerate its thermodynamical properties.
In section \ref{s:snd}, we study the dynamics of small fluctuations of density, for which we
use the Thomas-Fermi approximation, followed by a linearization procedure.
In section \ref{s:snde} we use the elasticity approach to sound propagation in logarithmic
condensate in the transverse harmonic trap.
Some discussion and conclusions are presented in section \ref{s:con}.

\scn{The model}{s:mod}
The logarithmic Bose-Einstein condensate is described by a condensate wavefunction
$\Psi = \Psi (\textbf{x},t)$,
which is normalized to a number $N$ of condensate particles of mass $m$:
\be\lb{e:norm}
\int_\vol |\Psi|^2 \dvol  = 
\int_\vol n \, \dvol = N
, \ee 
where  $n = \rho/m = |\Psi|^2$ being  particle density,
and which obeys a minimal $U(1)$-symmetric nonlinear Schr\"odinger equation
\ba
i \hbar \partial_t \Psi
&=&
\left[-\frac{\hbar^2}{2 m} 
\vena^2
+
V (\textbf{x})
- 
F (|\Psi|^{2})
\right]\Psi
,\nn\\ 
F (n) &\equiv&
b \ln{(n/n_c)}
,
\label{e:oF}
\ea
where 
$V (\textbf{x})$ is the external or trap potential,
$b$ is the logarithmic nonlinear coupling and $n_c$ is the critical density value
at which the logarithmic term turns off.
In the absence of  external potential, the ground state solution of \eq\eqref{e:oF}  
is a Gaussian function when coupling $b$ is positive \cite{ros68,bb76};
whereas
at negative values of $b$, the model allows topologically nontrivial solutions \cite{z19ijmpb}.

One can show that coupling 
$b$ is
linearly related to the wave-mechanical temperature $T_\Psi$,
defined as a thermodynamical conjugate
of the Everett-Hirschman entropy function.
The latter is quantum information entropy,
which emerges from the logarithmic term
when one averages the wave equation \eqref{e:oF}
in a Hilbert space
of wavefunctions $\Psi$.
In units where the Boltzmann constant is one,
it reads
$ 
S_\Psi 
= -\int_\vol 
|\Psi|^2 \ln{(|\Psi|^2/n_c)}
\, \dvol
$, 
further details can be found in \Ref \cite{az11}.
It is conjectured that $T_\Psi$
is
linearly related 
to the conventional (thermal) temperature $T$,
thus we can assume here that 
$ 
b  \sim T_\Psi \sim T
$, 
or
\be\lb{e:logtemp}
b  = \chi (T - T_c)
,\ee
where $T_c$ is a critical temperature 
(at which the logarithmic term switches off, similar to when density approaches $n_c$), and 
$\chi$ would be a scale constant, dimensionless in the chosen units.
This formula indicates that $b$ is not a fixed parameter of the model, 
instead its value
depends on the condensate's environment \cite{z18zna}.
The related phase structure is discussed in \Ref \cite{z19ijmpb}.

Furthermore,
in terms of particle density,
the
energy functional for the logarithmic condensate is given by
\be\lb{e:enfun}
E [n] =
\int_\vol
\left\{
\frac{\hbar^2}{2 m}
\left(\vena \sqrt n\right)^2
-
b n
\left[
\ln{(n/n_c)} -1
\right]
\right\}
\dvol
,
\ee
which
can be derived directly from \eq \eqref{e:oF}, cf. \Ref\cite{az11}.
From \eqs \eqref{e:oF} and \eqref{e:enfun}, 
one can see that the logarithmic condensate can switch between attractive and repulsive regimes as density's value
crosses $n_c$,
which leads to many profound effects only pertinent to the logarithmic Bose liquids.

%

To derive an expression for the chemical potential $\mu$,
let us recall that it is defined as the Lagrange multiplier 
that ensures constancy of the number $N$ under arbitrary variations of wavefunction.
Following a standard procedure \cite{psbook}, we use the stationary ansatz
$\Psi (\textbf{x},t) = \psi (\textbf{x}) \exp{(- i \mu t/\hbar)}$,
to obtain from \eqref{e:oF}:
\ba
\mu
=
- 
\frac{\hbar^2}{2 m} 
\frac{ \vena^2 
\psi}{\psi}
+
V (\textbf{x})
- 
b \ln{(|\psi|^2/n_c)}
,\label{e:chempot}
\ea
so that
for the free homogeneous condensate, we therefore obtain
\be\lb{e:cpcases}
\mu \approx
- b \ln{(|\psi|^2/n_c)} = - b \ln{(n/n_c)}
, 
\ee
where the approximation indicates that we have disregarded the kinetic energy term.

One can see that, similar to the attractive/repulsive behaviour discussed earlier, 
the 
logarithmic condensate's $\mu$ switches its sign as density
goes across the value $n_c$.
Thus, adding or removing particles to the logarithmic condensate is energetically favorable
in two of the four regions 
$\{n \gtrless n_c,\ T \gtrless T_c \}$,
and unfavorable in the others.
This, of course, has a crucial influence upon the stability of the free logarithmic condensate \cite{az11,z17zna,rc21}.

Furthermore, in the Madelung representation, the wave equation \eqref{e:oF} can be written in hydrodynamic form;
from which one can deduce the equation of state and related values \cite{z19mat}.
If we assume that $b$ does not contain the Planck constant in higher than the first power;
then
in the leading order approximation with respect to this constant,
we can neglect the second derivatives of density. 
This is a robust assumption,
as long as we do not consider
shock waves and other fluctuations with a non-smooth density profile.
Then 
the hydrodynamic equations acquire the perfect-fluid form.
Additionally assuming constant temperature, cf. formula \eqref{e:logtemp},
the equation of state $p=p(n)$ and speed of sound $c_s$ read:
\ba
p 
&=&
- 
\int\!
n F'(n)\, \drm n
+ {\cal O} (\hbar^2)
=
- b n + {\cal O} (\hbar^2)
,\lb{e:pcases}\\
c_s
&\equiv& 
\sqrt{\Der{p}{\dn}}
=
\sqrt{
-
\frac{n F'(n)}{m}
+ {\cal O} (\hbar^2)
}
\nn\\&=&
\sqrt{- \frac{b}{m}
+ {\cal O} (\hbar^2)
}
,
\lb{e:ccases}
\ea
where prime refers to an ordinary derivative,
and
${\cal O} (A)$ denotes any terms of order $A$;
for a sake of brevity,
we omit these from now on.
These formulae reveal that the logarithmic condensate is the ideal liquid, 
whose pressure is a linear function of density, and whose speed of sound does not depend on density.

\scn{Linear regime}{s:snd}
Let us consider the propagation of sound in a cloud or lump of logarithmic Bose-Einstein condensate
whose transverse
dimensions $R$ are significantly less than its axial dimension $H$ aligned along $z$ axis. 
In this setup, we assume that density fluctuations have a characteristic length scale
$\ell$ in the axial direction, $R \ll \ell \ll H$,
which is usually what happens in experiments 
\cite{aem95,akm97}. 
Along the lines of those experimental setups,
we expect that vertical variations of the confining potential 
are small compared to the characteristic length $\ell$;
also that the particles' density values are
sufficient for the Thomas-Fermi approximation to be
robust.

Since the problem becomes essentially one-dimensional,
the lump can be characterized by its velocity $v=v(z)$
and linear particle density
\be\lb{e:surf}
\sigma = \sigma (z) = \int n (\textbf{x})\, \drm x \drm y
,
\ee
where $(x,y)$ are coordinates in transverse dimensions.
Because, in the previous section, we established that the logarithmic \schrod~ equation
can be
approximately rewritten in
a hydrodynamic perfect-fluid form;
we can derive hydrodynamic equations for $v$ and $\sigma$
by integrating over the condensate lump's transverse coordinates. 
Following 
a standard procedure \cite{kp97},
we obtain from the continuity and Euler equations, respectively:
\ba
&&
\pDer{\sigma}{t} +
\pDer{(\sigma v) }{z} = 0
,\lb{e:cont}\\&&
m \sigma \Der{v}{t} +
\sigma \pDer{V}{z} = - \int \pDer{p}{z} \, \drm x \drm y 
= b \int \pDer{n}{z} \, \drm x \drm y
, ~~~~\lb{e:eucases}
\ea
where we used \eq \eqref{e:pcases} in the last step.
Here $V$ is a combined external potential due to the trap and electromagnetic interactions.
Using definition \eqref{e:surf}, the Euler equation \eqref{e:eucases} becomes
\be\lb{e:eucases2}
m \Der{v}{t} = -
\pDer{V}{z} 
+
\frac{b}{ \sigma} \pDer{\sigma}{z} 
=
-
\pDer{V_\text{eff}}{z}
, 
\ee
where 
$ 
V_\text{eff} = V - b \ln{(\sigma/\sigma_c)}
$ 
is the modified external potential.

Let us consider the linear approximation, while disregarding spatial variations of the external potential $V$.
In the linearized regime,
from \eqs \eqref{e:cont} and \eqref{e:eucases2} we obtain
the partial differential equation
\be\lb{e:wavcases}
\left(
\pDer{^2}{t^2} 
+
\frac{b}{m}  \pDer{^2}{z^2}   
\right)
\sigma
=0
,
\ee
which must be supplemented with the normalization condition
\be\lb{e:normlin}
\int\limits_0^H \sigma \, \drm z = N
,
\ee
which follows from \eqs \eqref{e:norm} and \eqref{e:surf}.

Equation \eqref{e:wavcases}
can be of either a hyperbolic or an elliptic type,
depending on the sign of logarithmic coupling.
Let us consider these cases separately.

\sscn{Case $b < 0$}{s:bneg}
If $b$ is negative, $b = - |b|$ (\textit{i.e.}, $\chi >0$ and $T < T_c$),
then \eq \eqref{e:wavcases} becomes a hyperbolic-type partial differential equation:
\be\lb{e:wavcasesneg}
\left(
\pDer{^2}{t^2} 
-
\bar c_s^2  \pDer{^2}{z^2} 
\right)  
\sigma
=0
,
\ee
where the velocity
$ 
\bar c_s
=
\sqrt{|b|/m}   
$ 
is similar to the previously derived formula \eqref{e:ccases}.
Its general solution is a superposition of righ- and left-traveling functions
\be\lb{e:sigsolneg}
\sigma (z,t) = 
G_R (z - \bar c_s t) + G_L (z + \bar c_s t)
.
\ee
If we supplement \eq \eqref{e:wavcasesneg} with the Cauchy initial conditions
\be
\sigma (z,0) = \sigma_0 (z)
, \ \
\pDer{}{t} \sigma (z,0) = j_0 (z)
, \ee
then solution \eqref{e:sigsolneg} 
can be rewritten
\be
\sigma (z,t) = 
\frac{1}{2}
\left[
\sigma_0 (z - \bar c_s t) + \sigma_0 (z + \bar c_s t)
\right]
+
\frac{1}{2 \bar c_s}
\int\limits_{z - \bar c_s t}^{z + \bar c_s t} j_0 (\zeta)\, \drm\zeta
,
\ee
which is known as the d'Alembert formula.
This solution can describe not only propagating but also standing waves,
because the latter can be regarded as a superposition
of left- and right-propagating modes.

The formulae above indicate that small density fluctuations propagate in
the direction perpendicular to the lump's cross section.
It also
reaffirms the one-dimensional character of sound propagation in the logarithmic Bose-Einstein condensate.

\sscn{Case $b > 0$}{s:bpos}
If $b$ is positive, $b = |b|$ (\textit{i.e.}, $\chi >0$ and $T > T_c$),
then \eq \eqref{e:wavcases} becomes an elliptic-type partial differential equation:
\be\lb{e:wavcasespos}
\left(
\pDer{^2}{t^2} 
+
\bar c_s^2  \pDer{^2}{z^2} 
\right)  
\sigma
=0
,
\ee
where the value
$ 
\bar c_s^2   
$ 
is defined as in \eq \eqref{e:wavcasesneg};
but in this case it is nor longer related to the velocity of acoustic oscillations in the system,
as we shall see below.

To construct its general solution,
let us begin by finding its basic solutions.
Using the separation of variables, \textit{i.e.}, assuming the ansatz
$\sigma (z,t) = Z(z) T (t)$, 
\eq \eqref{e:wavcasespos} splits into two ordinary second-order differential equations 
\be
T''(t)/T(t) = -K = - \bar c_s^2 Z''(z)/Z(z) 
,
\ee
where $K$ is an arbitrary constant.
Their general solutions can be easily found in terms of exponential functions,
hence we obtain
\bw
\be\lb{e:varsepsol}
\sigma (z,t) =
\left(
A_- \text{e}^{-(\sqrt{K}/\bar c_s) z}
+
A_+ \text{e}^{(\sqrt{K}/\bar c_s) z}
\right)
\left(
B_- \text{e}^{-\sqrt{-K} t}
+
B_+ \text{e}^{\sqrt{K} t}
\right)
,
\ee
where $A$'s and $B$'s are integration constants.
From this formula, one can immediately see that: depending on
whether $K$ is positive or negative, we have different types of solutions.
Let us consider them separately.\\

\textit{Branch}  $K > 0$.
In this case, we can redefine $K = \omega^2$, where $\omega$ is the frequency of the mode.
Then from \eq \eqref{e:varsepsol} we obtain basic solutions 
of \eq \eqref{e:wavcasespos}:
\be\lb{e:sigkpos}
\sigma_\omega (z,t) 
=
\left\{
\baa{ll}
\displaystyle
\left\{
\text{e}^{- \gamma z - i \omega t }
, \,
\text{e}^{- \gamma z + i \omega t }
, \,
\text{e}^{\gamma z - i \omega t}
, \,
\text{e}^{\gamma z + i \omega t}
\right\}
&\displaystyle
\ \text{if} \
z \in [0, H],\\ 
\displaystyle
\left\{
\text{e}^{- \gamma z - i \omega t}
, \,
\text{e}^{- \gamma z + i \omega t}
\right\}
&\displaystyle  
\ \text{if} \
z \in [0, +\infty) ,
\eaa
\right.
\ee
where $\gamma = \omega/\bar c_s = \omega \sqrt{m/|b|} $, 
and $H$ is assumed finite. 
In the bottom set of solutions, which are defined on the semi-infinite domain $z \in [0, +\infty)$,
we took into account the normalization condition \eqref{e:normlin},
which implies that density should vanish at spatial infinity.
Restricting ourselves to real values, one can write these solutions in the form
\be\lb{e:sigkpos2}
\sigma_\omega (z,t) 
=
\left\{
\baa{ll}
\displaystyle
\left\{
\text{e}^{- \gamma z} \cos{(\omega t )} 
, \,
\text{e}^{- \gamma z} \sin{(\omega t )}
, \,
\text{e}^{ \gamma z} \cos{(\omega t )}
, \,
\text{e}^{\gamma z} \sin{(\omega t )}
\right\}
&\displaystyle
\ \text{if} \
z \in [0, H],\\ 
\displaystyle
\left\{
\text{e}^{- \gamma z} \cos{(\omega t )} 
, \,
\text{e}^{- \gamma z} \sin{(\omega t )}
\right\}
&\displaystyle  
\ \text{if} \
z \in [0, +\infty) ,
\eaa
\right.
\ee
which reveals their physical meaning:
these solutions 
describe the temporal oscillations of density with frequency $\omega$
whose amplitudes exponentially decay or grow, as the coordinate $z$ increases.
Thus, in this case, the value $\bar c_s$ no longer plays the role of
the speed of sound,
but instead contributes to a spatial decay constant $\gamma$.

The general solution can be written as the Laplace transform
with respect to the eigenmodes \eqref{e:sigkpos}:
\ba
\sigma (z,t) 
&=&
\sum_n
\int\limits_{0}^{+\infty}
s_{n} (\omega)
\sigma_\omega^{(n)} (z,t)\, \drm \omega
=
\int\limits_{0}^{+\infty}
s_{1}(\omega)
\text{e}^{- (\omega/\bar c_s) z - i \omega t}\, \drm \omega
+
\int\limits_{0}^{+\infty}
s_{2}(\omega)
\text{e}^{- (\omega/\bar c_s) z + i \omega t }\, \drm \omega
+
\ldots
,
\ea
where the functions $s (\omega)$ are determined by the boundary or initial conditions,
and index $n$ enumerates modes given by \eqs \eqref{e:sigkpos} or \eqref{e:sigkpos2}. 
Note that, unlike the periodic Fourier transform; 
Laplace-transformed solutions are also valid 
in an infinite region of the space spanned by $z$.\\

\textit{Branch}  $K < 0$.
In this case, we can redefine $K = - \bar c_s^2 k^2$,
with $k$ being a $z$-component of the wavevector.
Then from \eq \eqref{e:varsepsol} we obtain basic solutions 
of \eq \eqref{e:wavcasespos}:
\be\lb{e:sigkneg}
\sigma_k (z,t) 
=
\left\{
\text{e}^{ - \Gamma t - i k z}
, \,
\text{e}^{ - \Gamma t + i k z}
, \,
\text{e}^{ \Gamma t - i k z}
, \,
\text{e}^{ \Gamma t + i k z}
\right\}
,
\ee
where $\Gamma = \bar c_s k = k \sqrt{|b|/m} $. 
Restricting ourselves to real values, one can write these solutions in the form
\be\lb{e:sigkneg2}
\sigma_k (z,t) 
=
\left\{
\text{e}^{- \Gamma t} \cos{(k z)} 
, \,
\text{e}^{- \Gamma t} \sin{(k z)}
, \,
\text{e}^{\Gamma t} \cos{(k z)}
, \,
\text{e}^{\Gamma t} \sin{(k z)}
\right\}
,
\ee
which reveals their physical meaning:
these solutions 
describe spatial oscillations of density with a wavenumber $k$,
whose amplitudes exponentially decay or grow with time.
Thus, in this case, the value $\bar c_s$ no longer plays a role of
the speed of sound
but contributes to the decay constant $\Gamma$.

The general solution can be written as the bilateral Laplace transform
with respect to the eigenmodes \eqref{e:sigkneg}:
\ba
\sigma (z,t) 
&=&
\sum_n
\int\limits_{-\infty}^{+\infty}
s_{n} (k)
\sigma_k^{(n)} (z,t)\, \drm k
=
\int\limits_{-\infty}^{+\infty}
s_{1}(k)
\text{e}^{ - \bar c_s k t - i k z}\, \drm k
+
\int\limits_{-\infty}^{+\infty}
s_{2}(k)
\text{e}^{ - \bar c_s k t + i k z}\, \drm k
+
\ldots
,
\ea
\ew
where the functions $s (k)$ are determined by the boundary or initial conditions,
and index $n$ enumerates modes given by \eqs \eqref{e:sigkneg} or \eqref{e:sigkneg2}. 

To summarize Sec. \ref{s:bpos}, in the linearized Thomas-Fermi approximation,
small density fluctuations in the logarithmic BEC model with positive $b$
are oscillations with an amplitude which varies in time or space.

\scn{Axial harmonic trap}{s:snde}
In the previous section, it was possible to disregard the trapping potential in the linear
regime, 
thus rendering the system effectively free.
In this section, we consider the propagation of sound
in the cigar-shaped logarithmic Bose-Einstein condensate,
which is confined in the transverse direction
using an external potential. 

Following previous section's arguments,
we regard this system as one-dimensional.
Moreover, we shall describe it as an elastic material,
whose total energy consists of the kinetic energy resulting from 
the condensate's bulk motion, elastic energy $E_\text{int}$ 
from particle interactions
and the contributions
from the trap potentials along $z$ and in transverse directions, $E_\text{trap}$. 
We shall assume that the confining potential density in the
transverse direction has a harmonic form
\be\lb{e:trpoten}
U_\text{trap} (\varrho) =
\frac{1}{2} m \omega^2 \varrho^2
, 
\ee
where $\varrho = \sqrt{x^2 + y^2}$ is the cross section's radial coordinate and $\omega$
the frequency of transverse oscillations of a particle of the condensate in
the trap. 

Assuming the cross section of the condensate's
lump to be locally uniform,
let us evaluate the elastic energy
$E_\text{el} = E_\text{trap} + E_\text{int}$.
The energy per unit length due
to the transverse trap is 
\ba
E_\text{trap} (z) 
&\equiv &
2 \pi \int U_\text{trap}(\varrho)\, n (\varrho,z) \varrho \,\drm\varrho
\nn\\&=&
\pi m \omega^2 
\int  n (\varrho,z) \varrho^3 \,\drm\varrho
, \lb{e:trelen}
\ea
where the integration is taken over the lump's cross section.
The interaction energy per unit volume reads
\be\lb{e:inten}
E_\text{int} (z) 
=
- 2 \pi b
\int  n (\varrho,z)
\left[
\ln{(n (\varrho,z)/n_c)}
-1
\right]
 \varrho\,\drm\varrho
, 
\ee
which
can be deduced from \eq \eqref{e:enfun}.

Furthermore,
for the harmonic trap,
the density in the Thomas-Fermi approach is given by
\be\lb{e:denstfz}
n (\varrho,z) =
n (0,z) 
\left(
1 - 
{\varrho^2}/{R^2}
\right)
,
\ee
where $R = R(z)$ is the lump's cross section radius \cite{bp96}.
Substituting this expression to \eq \eqref{e:surf},
we obtain 
number of particles per unit length
\be\lb{e:surfz}
\sigma (z)
=
2 \pi \int\limits_0^R 
n (\varrho,z) \varrho\, \drm\varrho =
\frac{1}{2} \pi n (0,z) R^2
,
\ee
while the trapping and interaction energies become
\ba
E_\text{trap} (z) 
&=&
\frac{1}{12}
\pi m \omega^2 
n (0,z) R^4
, \lb{e:trelenz}\\
E_\text{int} (z) 
&=&
\frac{3}{4}
\pi b\,
n (0,z) 
\left[
1-
\frac{2}{3}
\ln{\!\left(n (0,z)/n_c\right)}
\right]
R^2\!, ~~~~\lb{e:intenz}
\ea
according to \eqs \eqref{e:trelen} and \eqref{e:inten}, respectively.

Furthermore,
particle density on the trap's axis can be obtained 
as a solution of the equilibrium condition between 
the trapping and interaction energy.
We therefore obtain
\be\lb{e:noz}
n (0,z) = n_c
\exp{\!\left(
\frac
{3}{2}
-
\frac
{m \omega^2 R^2}{6 b}
\right)}
, \ee
from which
an expression  
for linear particle density immediately follows:
\be\lb{e:sigmR}
\sigma (z) =
\frac{1}{2}
\pi n_c R^2
\exp{\!\left(
\frac
{3}{2}
-
\frac
{m \omega^2 R^2}{6 b}
\right)}
,
\ee
with the use of \eq \eqref{e:surfz}.
One can regard \eq \eqref{e:sigmR} as
an equation for lump's radius as a function of the linear density.
Further studies depend on the sign of logarithmic coupling.
Similar to the previous section,
let us consider these cases separately.

\begin{figure}[htbt]
\begin{center}\epsfig{figure=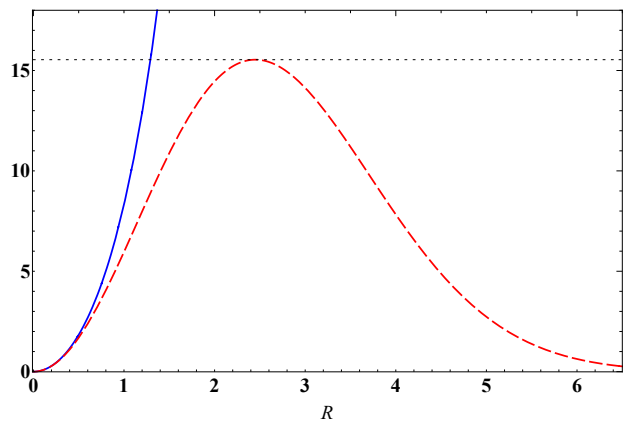,width=  \sc\columnwidth}\end{center}
\caption{
Linear density $\sigma(z)$ in units of $n_c \lambda_s^2$, versus the
radius of the lump
in units of  $\lambda_s$, for 
$b < 0$ (solid curve) and $b > 0$ (dashed).
The horizontal dotted line is drawn at $3\pi \sqrt\text{e}$
and represents $\sigma_\text{max}$. 
}
\label{f:sigmar}
\end{figure}

\sscn{Case $b < 0$}{s:ebneg}
Here we impose $b = - |b|$ so that \eqs \eqref{e:noz} and \eqref{e:sigmR} can be rewritten as
\ba
\sigma (z) &=&
\frac{1}{2}
\pi n (0,z) R^2
=
\frac{1}{2}
\pi n_c R^2
\exp{\!\left(
\frac
{3}{2}
+
\frac{R^2}{6 \lambda_s^2}
\right)}, ~~~\lb{e:sigmRneg}
\ea
where $\lambda_s = \sqrt{|b|/m}/\omega = \bar c_s/\omega$.
Thus, linear density in this case is a monotonously growing function of $R$, cf. a solid curve in Fig.\ref{f:sigmar}.

By inverting \eq \eqref{e:sigmRneg},
we obtain
\be\lb{e:Rsigmneg}
R (\sigma) =
\left\{
\baa{ll}
\displaystyle
\text{e}^{-3/4}
\sqrt{
\frac{2 \sigma}{\pi n_c} 
},&\displaystyle
\
 \omega = 0,\\ 
\displaystyle
\sqrt 6
\lambda_s
W^{1/2} 
\!\left(
\frac{\sigma}{3 \pi \text{e}^{3/2} n_c \lambda_s^2}  
\right)
,&\displaystyle  
\
\omega \not= 0 ,
\eaa
\right.
\ee
where $W (x)$ is the Lambert function.
For small values of the lump's radius
($R \ll \bar c_s/\omega$),
both $n (0,z)$ and $R$ vary as $\sigma^{1/2}$.

\begin{figure}[htbt]
\begin{center}\epsfig{figure=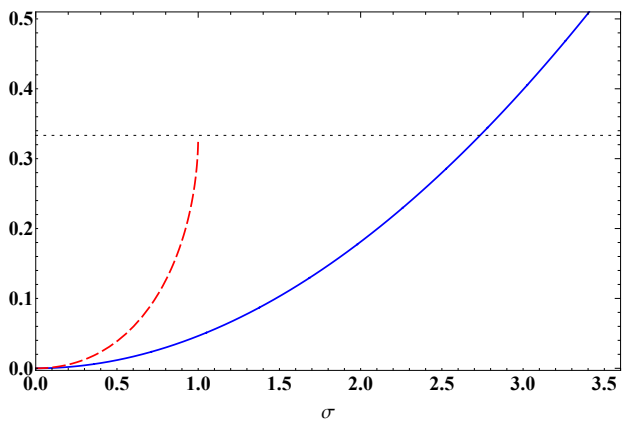,width=  \sc\columnwidth}\end{center}
\caption{
Elastic energy $E_\text{el}$
versus linear density $\sigma$, for 
$b < 0$ (solid curve) and $b > 0$ (dashed).
The units of energy and density are, respectively,
$m \omega^2 n_c \lambda_s^4$ and 
$n_c \lambda_s^2$ for $b < 0$,
and 
$E_\text{max}$ and $\sigma_\text{max}$
for $b > 0$.
The horizontal dotted line is drawn at $1/3$
and represents $E_\text{max}$. 
}
\label{f:elenr}
\end{figure}

Using \eqs \eqref{e:trelenz}, \eqref{e:intenz}, \eqref{e:sigmRneg}
and \eqref{e:Rsigmneg},
the elastic energy can be written as
\ba
E_\text{el} (z) 
=
\frac{1}{3}
m \omega^2 \sigma R^2
= 
2
m \omega^2 
\lambda_s^2
\sigma
W\!\left(
\frac{\sigma}{3 \pi \text{e}^{3/2} n_c \lambda_s^2}  
\right)\!, ~~\lb{e:elenzneg}
\ea
and plotted as a solid curve in Fig. \ref{f:elenr}.
Because the Lambert function increases at a slower rate than the linear function; 
the elastic energy increases less rapidly with growing $\sigma$ 
than it would in the case of confinement
by a rigid pipe with infinitely high walls (for which 
the potential energy $\sim \sigma^2$). 
This reflects the fact that
for a harmonic confining potential, an increase in the number
of particles per unit length results in expansion of the cloud
in the radial direction; 
thereby leading to a less rapid increase
of the total potential energy than would be the case for confinement
by rigid walls.

\sscn{Case $b > 0$}{s:ebpos}
Then $b =  |b|$ so that 
 \eqs \eqref{e:noz} and \eqref{e:sigmR} can be rewritten as
\be\lb{e:sigmRpos}
\sigma (z) =
\frac{1}{2}
\pi n (0,z) R^2
=
\frac{1}{2}
\pi n_c R^2
\exp{\!\left(
\frac
{3}{2}
-
\frac{R^2}{6 \lambda_s^2}
\right)}
,
\ee
being plotted as the dashed curve in Fig. \ref{f:sigmar}.
One can see that linear density in this case has an absolute maximum
\be
\sigma (z) \leqslant
\sigma_\text{max} = 3 \pi \sqrt{\text e} \, n_c \lambda_s^2
=
3 \pi \sqrt{\text e} \, n_c \bar c_s^2/\omega^2
,
\ee
at $R_\text{max} = \sqrt 6 \lambda_s$, cf. the dotted line in Fig.\ref{f:sigmar},
and then tends to zero as $R$ grows.
It is convenient to rewrite \eq \eqref{e:sigmRpos} in terms of the peak values:
\be\lb{e:sigmRpos2}
\sigma (z) =
\text e^{-1/2}
\sigma_\text{max} 
X^2
\exp{\!\left(
\frac
{3}{2}
-
X^2
\right)}
,
\ee
where we denoted $X = {R}/{R_\text{max}}$.

By inverting \eq \eqref{e:sigmRpos2},
we obtain
\ba
R (\sigma) =
\left\{
\baa{ll}
\displaystyle
\text{e}^{-3/4}
\sqrt{
\frac{2 \sigma}{\pi n_c} 
},&\displaystyle
\
 \omega = 0,\\ 
\displaystyle
- R_\text{max} 
W^{1/2}\!\left(-
\frac{\sigma}{\text{e} \, \sigma_\text{max}}  
\right)
,&\displaystyle  
\
\underset{\sigma \leqslant\sigma_\text{max} }{\omega \not= 0}
,
\eaa
\right.
\lb{e:Rsigmpos}
\ea
where
it should be remembered that $R (\sigma)$ is a two-valued function at $\omega \not= 0$.
Similar to the case of negative $b$,
for small values of lump's radius
($R \ll \bar c_s/\omega$),
both $n (0,z)$ and $R$ vary as $\sigma^{1/2}$.

Using \eqs \eqref{e:trelenz}, \eqref{e:intenz}, \eqref{e:sigmRpos2}
and \eqref{e:Rsigmpos},
the elastic energy can be written as
\ba
E_\text{el} (z) 
&=&
-
\frac{1}{3} 
\frac{\sigma}{\sigma_\text{max}} 
E_\text{max}
W\!\left(
- \frac{\sigma}{\text{e} \, \sigma_\text{max}}  
\right)
\leqslant
E_\text{max}
, ~~\lb{e:elenzpos}
\ea
where
$ 
E_\text{max} = m \omega^2 R_\text{max}^2 \sigma_\text{max}
=
18 \pi \sqrt{\text e}\, m \omega^2 n_c \lambda_s^4
$, 
and plotted by a dashed curve in Fig. \ref{f:elenr}.
Thus, the behaviour of linear density and elastic energy is similar to Sec. \ref{s:ebneg},
cf. Fig. \ref{f:elenr},
except that they are bound from above, by 
$\sigma_\text{max}$ and $E_\text{max}$, respectively.

\scn{Conclusion}{s:con}
In this paper, we studied  the propagation of 
small fluctuations of density in cigar-shaped logarithmic Bose-Einstein condensates, using the Thomas-Fermi and linear approximations.

It is shown that the dynamics of such fluctuations of the logarithmic condensate in a strongly anisotropic trap is effectively one-dimensional,
where the speed of sound is independent of particle density in a leading-order approximation with respect to the Planck constant.
This makes the logarithmic condensate similar to an ideal fluid in terms of classical averaged values.

In the linear regime of the Thomas-Fermi approximation,
one can disregard the influence of slowly varying trapping potentials,
thus describing the system as if it were free.
In that case,
small density fluctuations in the logarithmic BEC obey a second-order partial differential equation.
This equation can be of either a hyperbolic (d'Alembert-like) or elliptic (Laplace-like) type, 
for the negative and positive nonlinear coupling $b$, respectively.

Consequently, the behaviour of acoustic oscillations is different,
depending on the sign of coupling.
If it is negative, density fluctuations are waves propagating
without changing their initial shape and amplitude.
They can take the form of pulses (described by single propagating modes)
or standing waves (described by superpositions of modes).
On the other hand, 
small density fluctuations in the case of a positive value of $b$
are oscillations whose amplitude varies in time or space.

We then considered the case of logarithmic BEC in an axial harmonic trap. 
Using elasticity theory formalism, 
we studied the properties of the linear particle density and the elastic energy of density oscillations.
It turns out that
the elastic energy increases less rapidly with growing $\sigma$ 
than it would in the case of confinement
by a rigid pipe with infinitely high walls. 
This reflects the fact that
for a harmonic confining potential, an increase in the number
of particles per unit length results in the expansion of the condensate lump
in the radial direction; thereby leading to a less rapid increase
of the total potential energy than would be the case for confinement
by rigid walls.

Similar to the linear regime, in the trap,
density and energy behave differently, depending on the sign of nonlinear coupling.
If it is negative, their behaviour is qualitatively similar to the Gross-Piatevskii 
condensate: density is a monotonously growing function of lump's radius,
while energy is a monotonously growing function of density.
If the coupling is positive, the density is bound from above,
therefore elastic energy is defined on a finite domain of density:
it grows monotonously as a function of density until it reaches a global maximum.

It should be emphasized that the above-mentioned results were obtained assuming a number of approximations,
which do not necessarily hold in the presence of large density inhomogeneities.
The logarithmic condensate, being essentially nonlinear, is known to form such inhomogeneities,
because it behaves more like quantum liquid than a gas \cite{az11,z12eb,z19ijmpb}.
Apart from the nonlinear phenomena, shock waves and other objects with effectively non-smooth density profiles
can also violate the above-mentioned set of approximations.
The study of nonlinear and shock-wave effects in one-dimensional quantum Bose liquids under more general conditions 
will be the subject of future work.

\begin{acknowledgments}
This work is based on the research supported 
by the Department of Higher Education and Training of South Africa
and
in part by the National Research Foundation of South Africa (Grants Nos. 95965, 131604 and 132202).
Proofreading of the manuscript by P. Stannard is greatly appreciated.
\end{acknowledgments}



\def\PR{Physical Review}
\def\JMP{Journal of Mathematical Physics}
\def\CMP{Communications in Mathematical Physics}
\def\APNY{Annals of Physics}
\def\PL{Physics Letters}
\def\PSc{Physica Scripta}
\def\IJTP{Int. J. Theor. Phys.}
\def\GC{Gravitation \& Cosmology}
\def\APP{Acta Physica Polonica}
\def\ZN{Zeitschrift f\"ur Naturforschung}
\def\JPB{Journal of Physics B: Atomic, Molecular and Optical Physics}
\def\EPJ{European Physical Journal}
\def\EPL{Europhysics Letters (EPL)}
\def\GAFD{Geophysical \& Astrophysical Fluid Dynamics}
\def\ARMA{Archive for Rational Mechanics and Analysis}
\def\ARFM{Annual Review of Fluid Mechanics}
\def\JPCS{Journal of Physics: Conference Series}

\end{document}